\documentclass[conference]{IEEEtran}
\IEEEoverridecommandlockouts
\usepackage{cite}
\usepackage{amsmath,amssymb,amsfonts}
\usepackage{graphicx}
\usepackage{textcomp}
\def\BibTeX{{\rm B\kern-.05em{\sc i\kern-.025em b}\kern-.08em
    T\kern-.1667em\lower.7ex\hbox{E}\kern-.125emX}}

\usepackage[english]{babel}
\usepackage{graphicx}
\usepackage{hyperref}
\usepackage{url}
\usepackage{balance}
\usepackage[utf8]{inputenc}

\usepackage[table,usenames,dvipsnames]{xcolor}

\newcommand{\remarkP}[1]{{\color{Purple}{*Per: #1*}}}
\newcommand{\remarkE}[1]{{\color{Orange}{#1}}}
\newcommand{\remarkG}[1]{{\color{Green}{*Gita: #1}*}}

\renewcommand{\remarkP}[1]{}
\renewcommand{\remarkE}[1]{}
\renewcommand{\remarkG}[1]{}

\begin{document}

\title{Ethical AI-Powered Regression Test Selection}

\author{\IEEEauthorblockN{Per Erik Strandberg}
\IEEEauthorblockA{
\textit{Mälardalen University}, Sweden \\
per.erik.strandberg@mdh.se
}
\and
\IEEEauthorblockN{Mirgita Frasheri}
\IEEEauthorblockA{%
\textit{Aarhus University}, Denmark \\
mirgita.frasheri@ece.au.dk}
\and
\IEEEauthorblockN{Eduard Paul Enoiu}
\IEEEauthorblockA{%
\textit{Mälardalen University}, Sweden \\
eduard.paul.enoiu@mdh.se}
}

\maketitle

\begin{figure*}[ht]
  \begin{center}
    \includegraphics[width=0.94\linewidth]{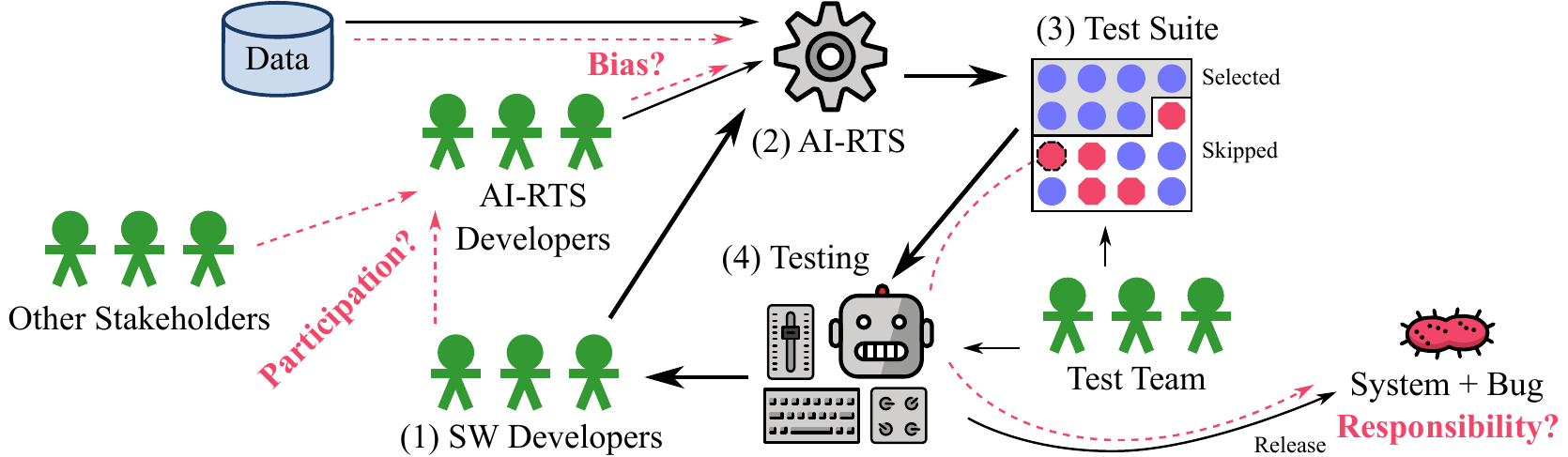}
    \caption{%
    Overview of a development and test cycle (thick arrows). 
    (1) Developers update the system under test.
    (2) An artificial intelligence powered regression test selection tool creates a test suite.
    (3) Because of resource constraints, only some test cases are executed in the system under test (4).
    It is possible, that a test case not selected for testing (highlighted with dashed border), could have detected a flaw that would otherwise lead to a bug in released software.
    Ethical challenges (bias, responsibility and participation) are highlighted with bold red text and question marks.
    (The figure is inspired by previous work on regression test selection and the flow of information in software testing \cite{strandberg2016, strandberg2019flow}.)
    }
    \label{fig:overview}
  \end{center}
\end{figure*}

\begin{abstract}
Test automation is common in software development;
often one tests repeatedly to identify regressions.
If the amount of test cases is large, one may select a subset and only use the most important test cases. The regression test selection (RTS) could be automated and enhanced with Artificial Intelligence (AI-RTS). This however could introduce ethical challenges. 
%
While such challenges in AI are in general well studied, there is a gap with respect to ethical AI-RTS. 
%
By exploring the literature and learning from our experiences of developing an industry AI-RTS tool, we contribute to the literature by identifying three challenges 
%
(assigning responsibility, bias in decision-making and lack of participation)
and three approaches (explicability, supervision and diversity).
%
Additionally, we provide a checklist for ethical AI-RTS to help guide the decision-making of the stakeholders involved in the process.
\end{abstract}


\section{Introduction}

A well studied problem in the software engineering domain is regression test selection (RTS), where one desires to select a good subset of tests from a larger pool of test cases, in order to ensure that a system under test (SUT) functions as expected.
Consequently, the automation of test selection represents a relevant research topic, for which several techniques have already been proposed, 
including artificial intelligence (AI) based methods.
An AI system can be defined as a decision-maker, capable of some autonomous actions, that operates in an environment, and tries to achieve a set of goals~\cite{legghutter2007collection}. 
The nature of these systems raises ethical concerns. While these are well studied ~\cite{jobin2019global} in general,  
no prior work has been written that considers the ethical aspects of RTS systems. 

In this paper we aim to start filling the gap in the existing body of knowledge on ethical AI-RTS by reasoning about a fictional AI-RTS tool, inspired by an actual tool implemented in 2014, \cite{strandberg2016}.
As opposed to an AI-RTS with just poor quality, an unethical AI-RTS may select poorly because of biased data, inscrutable decision-making, a poor understanding of what a just selection ought to be, or an implementation not anchored with stakeholder needs. 
We use a generic software development process with AI-RTS (as shown in Figure~\ref{fig:overview}) to illustrate different ethical challenges.

\section{Ethical Challenges in RTS}
\label{sec:challenges}
 

In early work on computer ethics, Moor reports of the invisibility factor:
the problem that one may not understand what the system is doing, possibly leading to abuse or bias (intentional or not)
\cite{moor1985computer}.
If an AI-RTS was non-transparent and hid reasons for decisions it would be difficult for developers or testers to take \textbf{responsibility}. 
Even if the RTS system would explain why some test cases are more important than others, if the operator cannot keep up with the many decisions made, the system as a whole can still be perceived as opaque. 


Cognitive \textbf{biases} are misconceptions that influence both humans and systems. One example is automation bias -- the tendency to have too much trust in automated suggestions or solutions. If a system is very reliable, then performance improves with automation. However, if a system has shortcomings, these tend to go unnoticed because humans disregard contradicting information \cite{cummings2017automation}. Imagine if an AI-RTS had a programming error where, instead of increasing priority of excluded tests after a week, it would only react after a month. One might argue that this would be an innocent programming bug and not bias. However, a consequence might be poorer overall performance of the testing process.

Suppose that an AI-RTS would be used at a company, but that there would be many parallel projects competing over the same resources for testing (this is already the case at companies such as Google \cite{memon2017taming}). The common and shared goal of ``selecting a good set of test cases for testing'' might expand to also include the task of distributing test resources between departments. 
On this larger scale there may be conflicting values and interests among stakeholders. In this case, we may have to think of a society-in-the-loop and not just a human-in-the-loop perspective. It may be motivated to oversee that a system respects fundamental rights, follows ethical values, 
by 
ensuring \textbf{participation} of various stakeholders, and by 
having humans watching the algorithms
\cite{hleg2019ai,rahwan2018society}

The most prevalent ethical principle in AI guidelines is transparency, and \textbf{explicability} is part of it \cite{jobin2019global}. 
Explicability means being able to explain as to why a certain decision has been made, which might be required for assigning responsibility. 
%
In order to understand decisions, a human-on-the-loop may need logs showing various internal states of the AI-RTS.
Deciding about the \emph{level of detail} of such logs, as well as \emph{what to log} 
is not trivial.
For the Therac-25 accidents, logging error messages was insufficient as too much output led to ``error message blindness,'' 
in particular because the operator could not trace errors to documentation \cite{leveson1993investigation}. 


As seen in Figure~\ref{fig:overview}, the subset of selected tests in the Test Suite is a result of the joint contribution of different sources, including the data used to train the decision-making AI, technical knowledge of developers, both AI-RTS and SW, that might not necessarily overlap, 
as well as empirical know-how gathered from experience on what usually works or not. 

One of several approaches for avoiding bias 
and improving justice and fairness of an AI \cite{jobin2019global},
is to conduct \textbf{supervision}.
However, most AI practitioners are unsure how to 
\cite{eitel2021beyond}.

Striving for \textbf{diversity} in AI is one way to improve participation,
and a wide range of guidelines promotes it, e.g.\ 
AI HLEG \cite{hleg2019ai}
and Amnesty
\cite{amnesty2018ai}. 
According to Jobin et al.\ \cite{jobin2019global} diversity can also target challenges with fairness and responsibility.

\section{Ethical Questions around AI-RTS}
\label{sec:conclusion}


In this paper, we discuss the challenges of responsibility, bias and participation. We argue that many approaches could tackle these ethical challenges and highlight the importance of explicability, supervision and diversity. 
In practice, these approaches are somewhat connected. E.g., supervision can help tackle bias, but so could diversity. According to Jobin et al.\ \cite{jobin2019global}, there are challenges when applying ethical guidelines for AI,
therefore, we suggest AI-RTS stakeholders consider the following questions: 
\begin{enumerate}

  \item How are humans in control of AI-RTS? 

  %
  %
  \item Can the system explain its test selection to a human?

  \item Are decisions and malfunctions suitably logged?
  
  \item Who is responsible if the 
  system under-performs?

  %
  %
  \item Have potential risks with the AI-RTS been considered? 
  
  \item What would be considered a just test selection?
 
  \item Is the impact of decisions (e.g.\ test coverage) monitored?
  
  %
  %
  \item Have all relevant stakeholders had the opportunity to participate in the design of the AI-RTS system? 
  
  \item How are stakeholders made aware of RTS decisions?
  
  \item What is the system's impact on human performance?

\end{enumerate}


Jobin et al.\ \cite{jobin2019global} have processed a large number of ethical guidelines to find common themes for AI in general. 
In particular, they identified that four challenges when applying ethical guidelines for AI are: how they ought to be interpreted, why they are important, where they are applicable, and how they should be implemented.
Of importance is also Eitel-Porter \cite{eitel2021beyond}, which mentions pitfalls in deploying ethical AI systems.
We would also like to point to Leveson, that reflects on safety-critical systems in a paper 30 years after her very influential original Therac-25 paper \cite{leveson2017therac, leveson1993investigation}.
Challenges she identified 
that are still relevant and might be applicable for AI-RTS include:
overconfidence in software,
unrealistic risk assessments,
as well as insufficient user and government oversight and standards.
Future work could include studies of already deployed AI-RTS tools, and explore, in retrospect, if the identified challenges are relevant.
Alternatively, future work could support the development of new AI-RTS tools.

%

\balance

\section*{Acknowledgments}
This research was partly funded by
the 20150277 ITS ESS-H grant, Innovation Foundation Denmark (AgroRobottiFleet),
the Poul Due Jensen Foundation (Centre for Digital Twin Technology at Aarhus University), and H2020 (957212).


\bibliographystyle{abbrv}
\bibliography{refs.bib}

\end{document}